\documentclass[%
 aip,
 rsi,%
 amsmath,amssymb,
 reprint,%
]{revtex4-2}

\usepackage{graphicx, color}
\usepackage{dcolumn}
\usepackage{bm}
\usepackage{array}
\usepackage{rotating}
\newcolumntype{P}[1]{>{\centering\arraybackslash}p{#1}}

\begin{document}

\title{Variable-temperature lightwave-driven scanning tunneling microscope with a compact, turn-key terahertz source}%

\newcommand{\ude}{\affiliation{Faculty of Physics and CENIDE, University of Duisburg-Essen, 47057 Duisburg, Germany}}

\author{Hüseyin Azazoglu}
\author{Philip Kapitza}
\author{Martin Mittendorff}
\author{Rolf Möller}
\author{Manuel Gruber} \email{manuel.gruber@uni-due.de} \ude

\date{\today}

\begin{abstract}
We report on a lightwave-driven scanning tunneling microscope based on a home-built microscope and a compact, commercial, and cost-effective terahertz-generation unit with a repetition rate of 100\,MHz.
The measurements are performed in ultrahigh vacuum at temperatures between 10\,K and 300\,K.
The cross-correlation of the pump and probe pulses indicate a temporal resolution on the order of a picosecond.
In terms of spatial resolution, CO molecules, step edges and atomically resolved terraces are readily observed in terahertz images, with sometimes better contrast than in the topographic and (DC) current channels.
The utilization of a compact, turn-key terahertz-generation system requires only limited experience with optics and terahertz generation, which may facilitate the deployment of the technique to further research groups.
\end{abstract}

\maketitle

\section{Introduction}

Scanning tunneling microscopy (STM) has, since its invention, been a valuable tool to image and probe the properties of individual atoms and molecules adsorbed on surfaces.
The transfer of energy and angular moment between the tunneling electrons and the investigated systems can trigger out-of-equilibrium dynamics, such as the vibration of a molecule \cite{stipe_single-molecule_1998}, the flipping of a spin \cite{heinrich_single-atom_2004}, and the switching of a molecule \cite{kim_action_2015}.
Because of the limited bandwidth of the transimpedance amplifiers, conventional STMs can only probe slow dynamics ($\geq 1$\,ms), while effects involving faster dynamics (e.g.\ spin flip) are measured in a steady-state regime.

To follow faster dynamics of nano-objects in the time domain, stroboscopic-based methods have been borrowed from the field of ultrafast laser spectroscopy.
The first implementation of pump-probe spectroscopy with STM dates from 1993 using photoconductive switches and a Au transmission line \cite{nunes_picosecond_1993}, which require particular samples.
Later designs used light (visible, mid-infrared, and infrared) focused on the STM junction to generate fast photocurrent pulses \cite{grafstrom_photoassisted_2002}.
However, the associated heating of the junction leads to thermal expansion/contraction issues, which are challenging to overcome \cite{tian_recent_2018}.

Thermal effects can be elegantly circumvented by resorting to voltage pulses as introduced by Loth \textit{et al}. \cite{loth_measurement_2010}, which was employed to investigate the spin dynamics of atoms \cite{yan_nonlocally_2017,paul_control_2017} and the charge dynamics of dopants in silicon \cite{rashidi_time-resolved_2016}.
The voltage pulses arriving at the STM junction are generally distorted because of the limited bandwidth and impedance matching of the wiring.
This limits the time resolution of the all-electronic pump-probe scheme, with a current record at 120\,ps (Ref.\citenum{saunus_versatile_2013}).

The coupling of terahertz (THz) pulses to the STM junction (hereafter referred to as lightwave-driven STM) provides a subpicosecond time resolution with negligible thermal effects \cite{cocker_ultrafast_2013,cocker_nanoscale_2021}.
A single-cycle THz pulse is generated and transmitted in free space to the STM junction, where the electric field of the THz-radiation is enhanced by a factor $10^5$ (Ref.~\citenum{kang_local_2009,peller_thesis_2020,peller_quantitative_2021}).
The rapidly varying electric field in the junction translates into a transient voltage between the tip and the sample, leading to a transient current. 
A THz-induced current $I_\text{THz}$, defined as the time-average of the transient current, is obtained when the transient voltage (of time-average zero) sweeps a non-linear region of the current-voltage characteristic at a repetition rate $f_\text{Rep}$.
For time resolved measurements, $I_\text{THz}$ is recorded for varying delays between two THz pulses (pump and probe pulses).
Further details may be found in recent reviews \cite{cocker_nanoscale_2021,tachizaki_progress_2021}.
This technique has experienced a growing interested and been used, for instance, to track the ultrafast motion of a pentacene molecule \cite{cocker_tracking_2016}, to investigate extreme tunnel currents through single atoms on a silicon surface \cite{jelic_ultrafast_2017}, to image graphene nanoribbons \cite{ammerman_lightwave-driven_2021} among other recent results \cite{yoshioka_tailoring_2018,yoshida_subcycle_2019,luo_nanoscale_2020,muller_phase-resolved_2020,peller_sub-cycle_2020,yoshida_terahertz_2021, abdo_variable_2021,sheng_launching_2022,wang_atomic-scale_2022,chen_single-molecule_2023}.
Besides THz table-top sources, it may be interesting to use intense narrow-band THz pulses from a free electron laser with STM, as it has been done for experiments based on atomic force microscopy \cite{kehr_near-field_2011}.

\begin{figure*}
\includegraphics{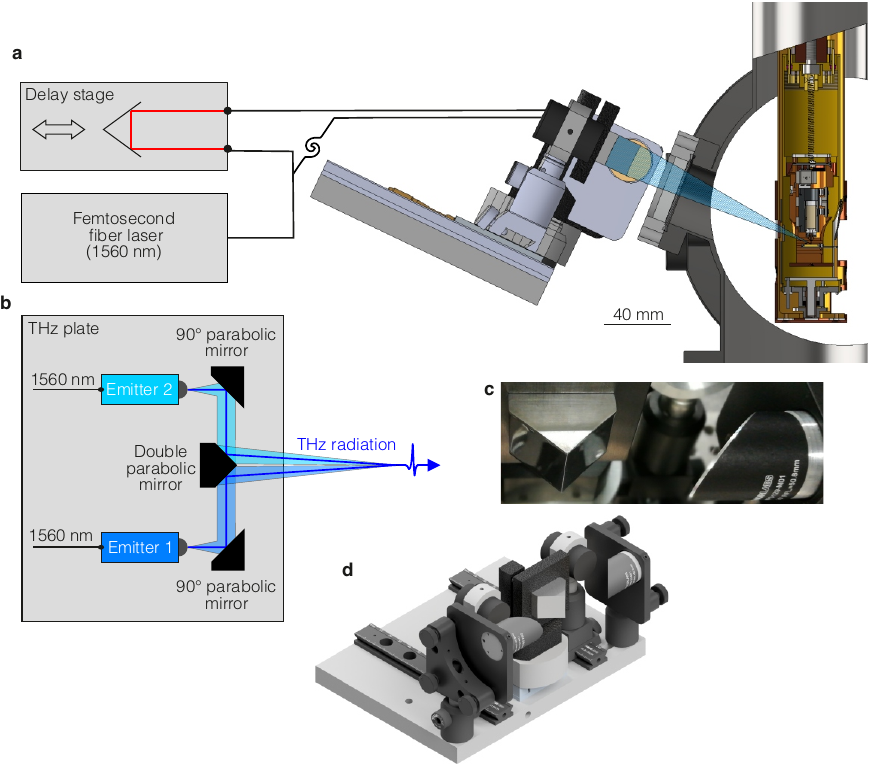}
\caption{\label{fig:overview} THz generation and coupling to the STM junction.
\textbf{a} Overview of the THz-STM setup composed of a femtosecond fiber laser (T-Light from Menlo systems, wavelength of 1560\,nm), a delay stage (adapted from Physik Instrumente allowing delays of up to 670\,ps), a ``THz plate'', and a home-built variable-temperature STM.
The angle between the THz beam and the surface normal is 70$^\circ$.
The solid black lines represent optical fibers for the transmission of the 1560\,nm-laser light, while solid red lines depict the laser beam in free space.
The delay stage is contained in a box, such that the user is effectively not exposed to beams in the free space.
\textbf{b} Detail of the THz plate including two photo-conductive THz emitters (Tera15-TX-FC from Menlosystems), two $90^\circ$\,off-axis parabolic mirrors, and a double off-axis parabolic mirror (home made).
The main role of this THz plate is to focus the THz radiations from two emitters into a single focal point, at the position of the STM junction.
\textbf{c}~Photograph of the double parabolic mirror along with one $90^\circ$\,off-axis parabolic mirror.
\textbf{d} Three-dimensional view of the THz plate.}
\end{figure*}

The reported setups are mostly based on Ti:Sapphire (800\,nm) or regenerative ytterbium-doped potassium gadolinium tungstate (Yb:KGW, 1030\,nm) lasers with output powers of 1\,W or larger typically providing pulses with energies between 1 and 40\,$\mu$J (see Table~\ref{tab:comparisonTHzSTM}). 
Tilted-pulse-front optical rectification in lithium niobate, the most commonly used for lightwave-drive STM (LDSTM), is particularly adapted for the generation of THz pulses with high electric-field amplitudes ($1$ to $16$\,kV/cm) provided that it is fed with high-energy optical pulses.
The large-amplitude THz beam may then be split to produce two pulses (pump and probe), which then need to be refocused on the tunnel junction.
Overall, the laser systems are relatively costly, the THz beam travels relatively large distances and is partially absorbed in air.
The THz (and optical) beam may then be placed in a dry-air purge box \cite{ammerman_lightwave-driven_2021,wang_atomic-scale_2022} to avoid the latest issue.
In combination to the inherent difficulties to operate an ultrahigh-vacuum STM, the LDSTM setups are fairly complex (albeit leading to exceptional results), which may discourage further adoption of the technique.
In the field of THz-spectroscopy, cost-effective, compact, turn-key instruments, for instance based on erbium fiber laser (1560\,nm, $~100$\,mW) widely spread for telecommunication applications, are commonly used.
Are such solutions compatible with LDSTM?

Here we report on a lightwave-driven STM instrument based on a home-built STM and a compact, commercial, and cost-effective THz-generation unit with a repetition rate of 100\,MHz.
The THz system is based on an erbium fiber laser connected with optical fibers to two photoconductive emitters.
The emitters and the focusing mirrors are mounted on a relatively compact plate ($20 \times 15$\,cm$^2$) placed in front of the ultrahigh-vacuum window of the STM at a distance of approximately 15\,cm from the STM junction.
Measurements at temperatures between 10 and 300\,K are realized.
The cross-correlation of the pump and probes pulses indicate a temporal resolution on the order of a picosecond.
In terms of spatial resolution, CO molecules, step edges and atomically resolved terraces are readily observed in THz images, with often better contrast than in the topographic and (DC) current channels.
The utilization of a compact, turn-key THz-generation system requires very limited experience with optics and THz generation, which may facilitate the deployment of the technique in further STM groups.

\section{Description of the setup}

The THz generation and the THz focusing components of our instrumentation are schematized in Figure~\ref{fig:overview}a.
A compact, turn-key, femtosecond fiber laser (T-Light from MenloSystems) delivers short pulses ($<90$\,fs) with a center wavelength of 1560\,nm at a repetition rate of 100\,MHz.
A fiber optic splitter splits the pulses into two fibers (0.2\,nJ per pulse after split), each of them connected with an optical fiber to a terrahertz photoconductive emitter (Tera15-TX-FC from Menlosystems) based on InGaAs/InAlAs heterostructures \cite{dietz_64_2013} providing an estimated output power of $\approx 30$\,$\mu$W.
One of the branch includes a delay line (adapted from Physik Instrumente, range of 670\,ps).
The transmission in optical fibers allows to separate the laser system and the delay stage from the STM frame, which would be more difficult for a free-space beam because of the vibrational damping of the STM frame.
The two emitters are mounted on a ``THz beams combiner plate'' (Figures~\ref{fig:overview}b,c) along with two off-axis parabolic mirrors, and a home made off-axis double parabolic mirror focusing the THz pulses at a distance of 120\,mm. 
The position of the THz plate is adjusted to have the focus point coinciding with the STM junction.
As the plate is mounted on the same frame as the STM, the vibrational damping of the STM frame does not affect the focus of the THz beam.
The generation and focusing of THz pulses is all realized on that plate, which is 20\,cm long and 15\,cm wide.
The femtosecond fiber laser and the delay stage are in a box of dimensions of $56 \times 46 \times 20$\,cm$^3$, making the THz-system relatively compact.

It should be mentioned that the usage of two THz emitters, for pump and probe pulses, is so far unique for LDSTMs.
This is particularly useful to reach higher THz output power (compared to one emitter).

\begin{figure}
\includegraphics{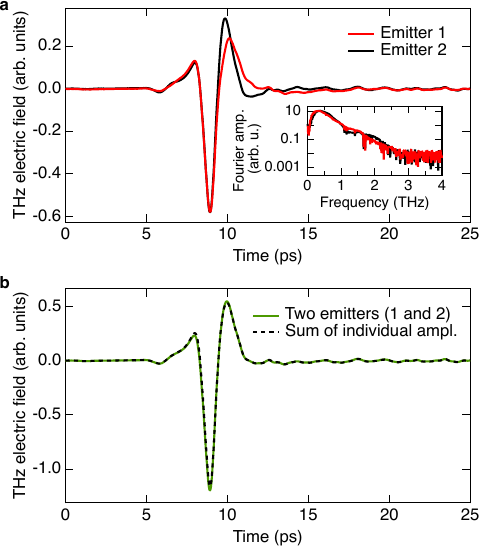}
\caption{\label{fig:thzPulseAir} Lineshapes of the THz pulses in the far field.
\textbf{a} Electric field transients at the THz focus point generated by Emitter 1 (red) and by Emitter 2 (black).
The corresponding Fourier transforms are shown in the inset, which exhibit a maximum at a frequency of approximately 0.3\,THz for Emitter 1 and 0.35\,THz for Emitter 2.
Emitter 2 is turned off during the data acquisition of Emitter 1, and \textit{vice versa}.
\textbf{b} THz transient (green) measured at the THz focus point with both emitters active (and no time delay between the generated pulses).
The dashed black line is the sum of the individual THz transients (red and black curves in \textbf{a}).
The data shown here are not deconvolved from the detector response, such that the actual pulse shape may deviate from these measurements.}
\end{figure}

The alignment of the parabolic mirrors is typically done on an optical table in ambient conditions.
A pin hole is placed at the focus point, and the THz radiation coming out from that pin hole is collimated with a lens toward a THz detector (Tera15-RX-FC).
That detector is triggered by the second output of the laser, which can be delayed using an internal delay line of the laser system.
This allows us to retrieve the waveform of the emitted THz pulses (Figure~\ref{fig:thzPulseAir}a).
As expected for electromagnetic waves in free space (without charge), the time integral of the measured signal over one cycle is approximately zero.
It should be noted that we do not exactly observe the waveform of the electric field, but the convolution of that waveform with the response function of the detector. 

The emitters produce relatively similar THz pulses (compare red and black curves in Figure~\ref{fig:thzPulseAir}a) with small variations at $\approx$10\,ps.
The Fourier spectra are maximum at frequencies of 0.3\,THz (Emitter 1) and of 0.35\,THz (Emitter 2).
The orientations of the emitters have been adjusted such that the polarizations of the THz electric fields are both vertical (leading to $p$ polarization upon coupling to the STM junction).
When both emitters are emitting, the total THz electric field at the focal point (green curve in Figure~\ref{fig:thzPulseAir}b with zero delay between the two pulses) corresponds to the sum of the individual components (dashed black curve).

\begin{figure}
\includegraphics{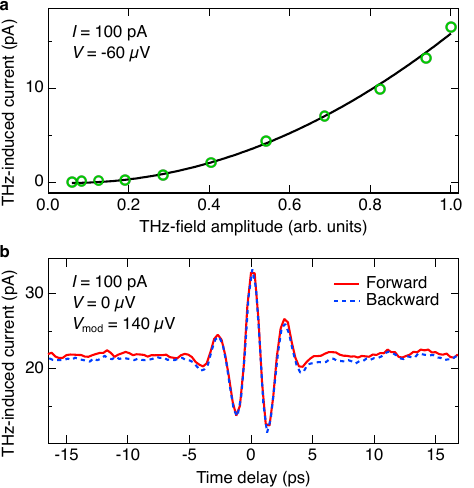}
\caption{\label{fig:pumpProbe} THz-induced tunneling current on a Ag(111) surface.
\textbf{a} Evolution of the average THz-induced current as a function of the amplitude of the incoming THz electric field (green circles).
The solid line is a quadratic fit to the data (linear term: $-1.9$, quadratic term: $17.7$).
\textbf{b} THz-induced current as a function of the time delay between the THz pump and probe pulses acquired for increasing (red) and decreasing 
(blue) delays.
A modulation voltage of 140\,$\mu$V at 970\,Hz was added (we verified that this modulation does not affect the detection of the THz-induced current).
The duration of the pulses is on the order of 1\,ps, which determines the temporal resolution.
Further field-sensitive analysis of the pulse shape can lead to an improved temporal resolution.
Measurements in \textbf{a} and \textbf{b} were performed at room temperature with the feedback current loop active (regulating on the absolute value of the current).}
\end{figure}

The coupling of the THz pulse to the STM junction leads to a transient voltage $v_\text{THz} (t)$, whose time integral is zero over one cycle.
The induced tunneling current is $i_\text{THz}(t) = G\left( v_\text{THz} (t) \right)$, where $G$ is the voltage-dependent conductance of the STM junction.
Because of the limited bandwidth of the current amplifier of the STM (cutoff frequency of $\approx 1.5$\,kHz), we are only sensitive to the time-average of $i_\text{THz}$
\begin{eqnarray}
I_\text{THz} =& f_\text{Rep} \int_0^{1/f_\text{Rep}} i_\text{THz}(t) \cdot dt \nonumber \\
=& f_\text{Rep} \int_0^{1/f_\text{Rep}} G\left( v_\text{THz} (t) \right), \label{eq:ITHz}
\end{eqnarray}
where $f_\text{Rep}$ is the repetition rate of the THz pulse.
$I_\text{THz}$ is referred to as THz-induced current.
As the time-average of $v_\text{THz} (t)$ is zero, the conductance $G$ must be non-linear with the voltage to obtain a finite THz-induced current.
Said differently, a non-linear $G$ is required to rectify the THz-induced voltage transient.

It should be noted that the transient current $i_\text{THz}$ has a non-zero value for only $\approx 10$\,ps such that the time integral in Eq.~\ref{eq:ITHz} is independent of the repetition rate ($1/f_\text{Rep}$ is typically much larger than one ns).
$I_\text{THz}$ then scales linearly with the repetition rate and indirectly depends on the amplitude of the transient voltage.

$I_\text{THz}$ is typically small and only represents a small fraction of the total tunneling current.
It is therefore measured by lock-in detection by chopping the THz beam at $\approx$1\,kHz.
The chopping is effectively done by modulating the voltage of the emitter electrodes (on/off with duty cycle of 50\,\%).
The fine alignment of the THz beam to the STM junction is done by maximizing the THz-induced current.

Figure~\ref{fig:pumpProbe}a shows the evolution of the THz-induced current as a function of the amplitude of the THz electric field (characterized in free space) measured on a Ag(111) surface.
The evolution is highly non linear (mainly quadratic), as expected, because we are sensitive to the rectified current, which depends on the details of the current-voltage characteristics.
Because of the relatively linear current-voltage characteristics of Ag(111) at low voltages, we are presumably only rectifying a small fraction of the induced current.
The tip-sample distance needs to be extremely small (resistance on the order of 1\,M$\Omega$) to have a measurable signal (exponential dependence on the tip-sample distance as the transient voltage is expected to be independent of the tip-sample distance in the tunneling regime\cite{peller_quantitative_2021}).
A typical DC voltage of 1\,V would lead to a large current on the order of 1\,$\mu$A that would cause issues with stability and surface degradation.
We therefore use relatively low voltages of $\approx 100\,\mu$V.
Under these conditions, the THz-induced current represents $\approx 10$\,\% of the total current, which is readily extracted by the lock in.

With the repetition rate of our system ($f_\text{Rep} = 100$\,MHz), a THz-induced current of 10\,pA corresponds to an average of 0.6\,electrons rectified per THz pulse.
We are therefore in a regime of one to few electrons per THz pulse on Ag(111) with our instrumentation.

In order to determine the time resolution of the instrument, we use a pair of THz pulses with variable delay between them.
The obtained cross-correlation function for increasing (decreasing) delays is shown in red (blue) in Figure~\ref{fig:pumpProbe}b.
A maximum THz-induced current of 33\,pA is obtained for zero time delay, and a minimum of 12\,pA for a delay of 1.2\,ps.
This 1.2\,ps represents a higher bound of the time resolution of the system.

\section{Experimental results}

\begin{figure}
\includegraphics{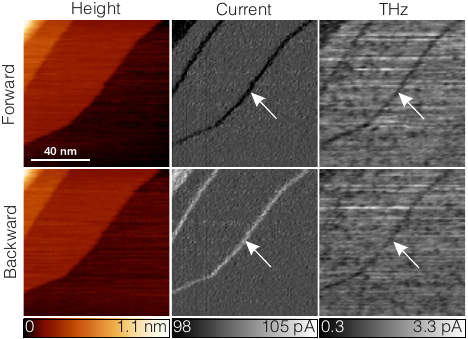}
\caption{\label{fig:stepEdge} THz-induced current at step edges.
Apparent height (left), average current (middle), and THz-induced current (right) images of a Ag(111) surface for left-to-right (top) and right-to-left (bottom) scan directions.
Because of the limited bandwidth of the feedback-loop, a height increase (decrease) in the scan direction causes a temporary increase (decrease) of the current (on the order of 1\,\%).
In the THz-induced current images, the step edge appear as a depression (the signal decreases by $\approx 25$\,\%), independently of the scan direction. 
The arrows point at the same step edge in different images.
The images have been acquired at room temperature with $I_\text{DC} = 100$\,pA and $V_{DC} = 250$\,$\mu$V at a scanning speed of 24.5\,nm/s.
These and the following STM images were processed with WSxM \cite{horcas_wsxm_2007}.}
\end{figure}

We first focus on large images of Ag(111) where the relative height of the tip, the current, and the THz-induced current were simultaneously recorded (Figure~\ref{fig:stepEdge}).
The step edges are observed as darker lines in the THz channel (see arrow in the top right image of Figure~\ref{fig:stepEdge}).
A similar behavior is observed in the current channel (top center), which is caused by the slow response of the feedback current loop and the tip-sample distance being too large for a short moment.
For the backward image (tip scanning from right to left), the slow response of the feedback causes an increase of the current (bottom center) due to a temporary too short tip-sample distance.
In contrast, the step edges remain darker in the THz image.
The darker lines in the THz image, although being sensitive (in intensity) to the tip-sample distance, do not exhibit a change of contrast.
These lines are therefore not directly related to the current.
We propose that the $I-V$ characteristic, and in particular $d^2 I / dV^2 (V)$, of the step edge slightly differ from that of the terrace \cite{li_local_1997} leading to a different rectification of the THz pulses.

\begin{figure}
\includegraphics{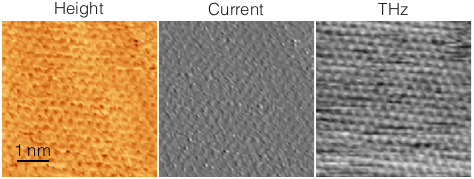}
\caption{\label{fig:atomicResolution} THz-induced current on a Ag(111) terrace.
Apparent height (left), average current (middle), and THz-induced current (right) images of a Ag(111) surface with atomic resolution.
The images have been acquired at RT with $I_\text{DC} = 100$\,pA and $V_{DC} = -70$\,$\mu$V.}
\end{figure}

The spatial resolution of the LDSTM is probably better visible in Figure~\ref{fig:atomicResolution}.
The atoms of the surface, slightly discernible in the topographic image, can be better observed in the THz image (raw data).
This data shows the atomic-resolution capability of the technique and of the instrument.

\begin{figure}
\includegraphics{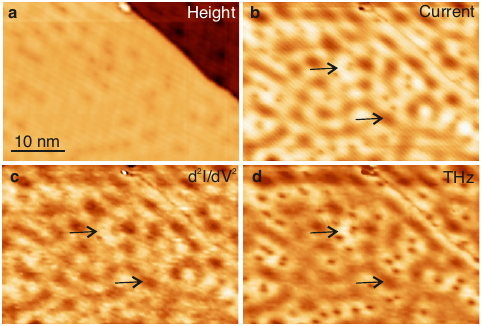}
\caption{\label{fig:CO_Ag111} Impurities, presumably CO molecules, on a Ag(111) surface.
\textbf{a} Apparent height, \textbf{b} average current, \textbf{c} $d^2I / dV^2$, and \textbf{d} THz-induced current images of a Ag(111) surface imaged at 80\,K and $V_{DC} = 0$\,V ($I_\text{DC} = 85$\,pA).
A modulation voltage of 280\,$\mu$V at 977\,Hz was applied and $d^2I / dV^2$ was recorded with a lock in.
The current feedback loop is fed with the absolute value of the current.
}
\end{figure}

To evidence the correlation between the rectification of the tunneling voltage by the non-linear current-voltage characteristic and the THz-induced current, we performed an experiment using a purely AC-tunneling voltage with an amplitude of 280\,$\mu$V without DC-component.
Instead of working with the tunneling current directly, the current feedback loop is fed with the absolute value of the current.
The observed average tunneling current results from the rectification of the voltage modulation by the tunneling junction.
Figure~\ref{fig:CO_Ag111} shows the apparent height, average current, $d^2I /dV^2$, and the THz-induced current acquired at 80\,K on a Ag(111) surface.
The surface exhibits $\approx 1$\,nm wide defects (observed as depressions in Figure~\ref{fig:CO_Ag111}a), which presumably are CO molecules\cite{kulawik_interaction_2005}.
As $V_{DC} = 0$, the measured average current is sensitive to the non-linearity of the current-voltage characteristic, which rectifies the sinusoidal voltage modulation.
To illustrate that, we acquired an image of the second harmonic of the current with a lock in, which corresponds to $dI^2/dV^2 (V = 0)$ (Figure~\ref{fig:CO_Ag111}c).
The two images are very similar.

The THz-induced current (Figure~\ref{fig:CO_Ag111}d) is also the result of a voltage rectification, this time not of the modulation voltage but of the repeated voltage transients induced by the THz pulses.
The finite $dI^2/dV^2$ at $V_{DC} = 0$ largely contributes to the rectification of the voltage transient as the modulations of the $dI^2/dV^2$ signal are visible in the THz image (see for instance arrows in Figure~\ref{fig:CO_Ag111} showing areas of larger and smaller intensities).
The remaining features of the THz image, for instance the well-resolved impurities, are likely originating from higher-order derivatives and from the larger voltage amplitude of the transient.
For completeness, we mention that the backward scan images look identical (not shown).

It appears that the small non-linearity of the current-voltage characteristic induced by surface states and impurities rectify a (presumably small) fraction of the THz-voltage transients leading to THz images.
We carried additional work on non-linear current-voltage characteristics at larger voltages, which indicate THz-induced voltage amplitude on the order of a few hundreds millivolt. 
The waveform of the voltage transient remains to be better characterized in future studies.

\section{Conclusion}
We have developed an ultrahigh-vacuum variable-temperature LDSTM using a compact, commercial, and turn-key THz-generation system based on a cost-effective femtosecond erbium fiber laser.
The instrument has a time resolution on the order of one picosecond and a sub-nanometer spatial resolution providing atomically-resolved THz images of a Ag(111) surface.
The proposed design requires a limited experience with optics and the generation of terahertz pulses, which facilitate the deployment of the technique.

\begin{acknowledgments}
We thank Tobias Roos, Detlef Utzat, and Doris Tarasevitch for technical support. 
Funding from the Deutsche Forschungsgemeinschaft (DFG; Project-ID 278162697 - CRC 1242, Projects A04, A08, and B09) is acknowledge.
\end{acknowledgments}

\section*{Conflict of Interest}
The authors have no conflicts to disclose.

\section*{Data Availability}
The data that support the findings of this study are available from the corresponding author upon reasonable request

\begin{sidewaystable}
\caption{\label{tab:comparisonTHzSTM} LDSTMs with key parameters related to the generation of the THz pulses. UDE and PCA stand for University of Duisburg-Essen and photoconductive antenna, respectively.}
\begin{ruledtabular}
\begin{tabular}{lP{2cm}P{2cm}P{2cm}P{2cm}P{2cm}P{2cm}P{2cm}P{2cm}P{2cm}}
Location&University of Alberta \cite{cocker_ultrafast_2013} &University of Regensburg \cite{cocker_tracking_2016}&Yokohama National University \cite{yoshioka_tailoring_2018} & University of Tsukuba \cite{yoshida_subcycle_2019} &Fritz Haber Institute \cite{muller_phase-resolved_2020}&University of Stuttgart \cite{abdo_variable_2021}&Michigan State University \citep{ammerman_lightwave-driven_2021}&University of California \cite{wang_atomic-scale_2022} &UDE\\
\hline
Publication year & 2013 & 2016 & 2018 & 2019 & 2020 & 2021 & 2021 & 2022 & 2023 \\
Laser wavelength (nm) & 800 & 1030 & 800 & 1035 & 800 & 1032 & 1030 & 820 & 1560\\
Energy per optical pulse ($\mu$J) & 4 & 16 & ? & 0.8 to 40 & 3 & 0.5 to 40 & 18 & $1.6 \times 10^{-3}$ & $0.2 \times 10^{-3}$\\
Repetition rate (MHz) & 0.25 & 0.61 & 0.001 & 1 to 50 & 1 & 0.5 to 41 & 1 & 1000 & 100\\
THz source & Large area PCA & Lithium niobate  & Lithium niobate &  Lithium niobate & Spintronic terahertz emitter & Lithium niobate & Lithium niobate & Plasmonic PCA & PCA\\
Splitted beam & THz & THz & THz  & &  & optic & THz & optic & optic\\
\end{tabular}
\end{ruledtabular}
\end{sidewaystable}

\bibliography{thzInst}

\end{document}